\documentclass[structabstract]{aa}
\usepackage[dvips]{graphicx}
\usepackage{txfonts}
\usepackage{color}
\usepackage{natbib}
\bibpunct{(}{)}{;}{a}{}{,} 
\usepackage{psfrag}
\usepackage{version}

\begin{document}


\title{Heteroatom-doped hydrogenated amorphous carbons, a-C:H:$X$}
\subtitle{
`volatile' silicon, sulphur and nitrogen depletion, blue photoluminescence, diffuse interstellar bands and ferro-magnetic carbon grain connections}
 
\author{A.P. Jones}
           
    \institute{CNRS, Institut d'Astrophysique Spatiale, UMR8617, Orsay F-91405, France\\[-0.3cm]
         \and  Universit\'e Paris Sud, Institut d'Astrophysique Spatiale, UMR8617, Orsay F-91405, France\\[0.1cm]
    \email{Anthony.Jones@ias.u-psud.fr} }

    \date{Received 12 April 2013 / Accepted 22 May 2013}

   \abstract
{Hydrogenated amorphous carbons, a-C:H, can incorporate a variety of heteroatoms, which can lead to interesting effects.}
{To investigate the doping of interstellar a-C:H grains with, principally, Si, O, N and S atoms within the astrophysical context.}
{A search of the literature on doped a-C:H reveals a number of interesting phenomena of relevance to astrophysics.}
{$X$ dopants in a-C:H:$X$ materials can affect the $sp^3$/$sp^2$ ratio ($X=$ Si, O and N), lead to blue photoluminescence (undoped or $X=$ N), induce ferromagnetic-like behaviour ($X=$ N and S) or simply be incorporated (depleted) into the structure ($X=$ Si, O, N and S). Si and N atoms could also incorporate into fullerenes, possibly forming colour-centres that could mimic diffuse interstellar bands.}
{Doped a-C:H grains could explain several dust-related conundrums, such as: `volatile' Si in photo-dissociation regions, S and N depletion in molecular clouds, blue luminescence, some diffuse interstellar bands and ferromagnetism in carbonaceous grains. }


\maketitle

\section{Introduction}

This work takes a look at the incorporation, or doping, of Si, O, N, S, Mg and Fe heteroatoms into amorphous hydrocarbon (a-C:H) grains and grain mantles in the interstellar medium (ISM) and in circumstellar regions. The aim is to search for common ground in some of the currently unresolved issues, such as: where and how does a-C:H grain formation and mantle accretion occur, and what is the composition and structure of the material? 
Given the current interest in carbonaceous grain (re-)formation and mantle accretion \citep[{\it e.g.},][]{2011A&A...530A..44J} and its important role in dust evolution it seems timely to re-consider some of the fundamentals of accretion and dust formation. 

Issues that appear to be of relevance and that might help to shed some light on the mechanics of grain (re-)formtion and accretion are: 
why is of the order of 10\% of Si not incorporated silicates,  
why is $\simeq 10-40$\% of Si apparently liberated from dust in photon-dominated regions (PDRs), reflection nebulae and stellar jets,   
can N be sequestered into dust, and 
why does S suddenly disappear from the scene in dense clouds?

The relatively high observed gas phase abundance of Si in diffuse clouds has been known of for a long time \cite[{\it e.g.},][]{1996ARA&A..34..279S}. Further, some $10-30$\% of Si is to be found in the gas phase in the 
\object{OMC\,1}, \object{S\,171}, \object{$\rho$\,Oph} and \object{$\delta$\,Sco}  PDRs \citep{2000A&A...356..705R,2003A&A...412..199O,2006ApJ...640..383O},  $20-30$\% of Si in the gas in the reflection nebula \object{NGC\,7023}  \citep{2000A&A...354.1053F}, $\gtrsim 10-40$\% of Si in the gas in the form of SiO in the \object{HH212} protostellar  jet \citep{2012A&A...548L...2C} and $\simeq 50$\% of Si in the gas in the \object{G333.6-0.2} H{\footnotesize II} region  \citep{1993ApJ...413..237C}.  These numbers are similar to the release of Si into the gas phase, via erosion from dust, in violently-shocked regions  \citep{2002ApJ...579..304W,2006A&A...456..189P,2009SSRv..143..311S}.  It is something of a conundrum as to why comparatively-benign PDR regions appear to be equally as destructive for Si-containing dust as supernova-generated shocks? Further, the observed SiO abundances in Herbig Haro jets appear to pose a challenge for current models dust erosion in shocks \citep{2012A&A...548L...2C}.  

The apparent `volatility' of Si in the ISM has been studied by \cite{1998ApJ...499..267T} and \cite{2000JGR...10510257J} and it was found that Si is preferentially released from dust before either Mg or Fe, contrary to the expectations for incident-ion sputtering \citep{2000JGR...10510257J}, and that the Si appears to be locked in a phase with a binding energy of $\approx 2$\,eV \citep{1998ApJ...499..267T}.

\section{a-C(:H) grain formation and mantle accretion}
\label{sect_aCH_accretion}

It has been proposed that the accretion of C and H atoms in the diffuse ISM leads to the formation of a-C:H mantles on dust in the transition from diffuse to the dense molecular clouds and that this process is consistent with many observables and with the evolution of interstellar dust \citep[{\it e.g.},][]{1990QJRAS..31..567J,2012A&A...540A...1J,2012A&A...540A...2J,2012A&A...542A...98J,2013A&A...xxx...xxx}. This scenario is consistent with the observation that the diffuse ISM hydrocarbon dust appears to be just that, {\it i.e.}, its composition is dominated by C and H atoms and there is little or no evidence for O or N heteroatoms being incorporated within the carbonaceous grains \citep[{\it e.g.},][]{2002ApJS..138...75P,2005A&A...432..895D}. However, in the denser ISM the remnant gas phase carbon will be mopped up and other abundant gas phase atomic (O, N, Si, \ldots) or radical/ionic species (CH$^{(+)}$, NH, CO, HCO, \ldots) will be incorporated into the accreting CH-dominated mantles. Similarly, in the dust forming regions around evolved stars heteroatoms can be incorporated into carbonaceous dust directly or through the incorporation of polyatomic radical or molecular species into the structure. Thus, the newly-formed a-C:H dust and a-C:H mantles can be doped with other elements. In the following sections we explore the effects of heteroatom incorporation into a-C:H(:$X$) materials, where $X=$ Si, O, N, S, Mg or Fe.

\section{Si-doping}
\label{sect_Si_doping}

Si can incorporate into the network of a-C:H materials where it promotes the formation of $sp^3$ carbon clusters \citep[{\it e.g.},][]{2004PhRvB..69d5410Y,2004ApPhL..85.4022R}. Conversely, amorphous silicates can be doped with hydrocarbons to yield  Si(CH$_3$)$_{1,2}$ structures \citep{2002ApJ...568L.131G}. Si can therefore be incorporated into an a-C:H $sp^3$ phase during dust formation around evolved stars or during carbonaceous mantle accretion (at the few atomic percent level\footnote{Given that $\approx 10$\% ($\cong 3$\,ppm) of Si \citep{1996ARA&A..34..279S} and $140\pm20$\,ppm of carbon \citep{1996ApJ...467..334C} in the gas phase in the diffuse ISM.}) in regions with $1 \lesssim A_V \lesssim 3$.  During accretion in the ISM, at low $A_V$, UV photons will tend to promote aromatic a-C formation and prevent aliphatic-rich a-C:H formation and, thus, hinder Si incorporation into the accreting a-C:H mantles. For $A_V \gtrsim 3$ ice mantles begin to form and a-C:H accretion will probably already have soaked-up all of the available gas phase carbon by this point. Conversely, in dust-forming circumstellar shells, any Si incorporated into a-C:H will be liberated during the aliphatic-to-aromatic ($sp^3 \rightarrow sp^2$) transformation when the dust encounters the harder ambient interstellar radiation field (ISRF). 

We note that IR absorption of Si$-$H bonds in $4.4 - 4.7\,\mu$m region is not observed in the diffuse ISM and that the IR signatures of any Si$-$O bonds within an a-C:H:Si:O material will be masked by the much stronger silicate features. This, then  leaves  Si$-$Si and Si$-$C bonds as tracers of the Si incorporated into any a-C:H:Si material. Given that the diffuse ISM is much richer in gas phase C ($\simeq 140$\,ppm) than Si ($\simeq 3$\,ppm), it is statistically most likely that the incorporated Si atoms will be bound to C atoms. Thus, a search should be made for the IR absorption of Si$-$C bonds in the $11\,\mu$m region, which would be analogous to the $11.3\,\mu$m feature in solid SiC. However, any observable feature\footnote{If it is not completely masked by the strong $\simeq 10\,\mu$m silicate absorption band.} will probably occur at longer wavelengths because of the Si atom incorporation into a much less refractory amorphous hydrocarbon phase than solid SiC. Additionally, Si(CH$_3$)$_2$ structures will produce bands at $7.95$ and $12.7\,\mu$m \citep{2002ApJ...568L.131G}. 

The energies for the most likely Si-containing bonds are shown in Table~\ref{table_Si_bondE}. As discussed above, the most likely signature for Si-doped a-C:H will be the Si$-$C bond, with a bond energy of $\sim 3.1$\,eV, consistent with the conclusion of \cite{1998ApJ...499..267T} that a significant fraction Si is locked into a relatively volatile dust material with a binding energy of $\simeq 2$\,eV. 

\begin{table}
\caption{Typical Si$-X$ and C$-X$ bond energies [ eV ].}
\begin{center}
\begin{tabular}{lclc}
                                  & &                            &                          \\[-0.35cm]
\hline
\hline
                                  & &                             &                          \\[-0.35cm]
    Bond                 & bond energy  &   Bond                 & bond energy    \\[0.05cm]
\hline
                                &   & &                    \\[-0.25cm]
    Si$-$Si                &      2.34    &  C$-$S                &    2.82               \\
    Si$-$C                &       3.12    &  C$-$Si               &    3.12         \\
    Si$-$H                &       3.35    &  C$-$N                &   3.16         \\
    Si$-$O                &       3.81    &  C$-$C                &   3.59        \\
    & &   C$-$O                &   3.71             \\[0.05cm]
    & &   C$=$S               &    5.94            \\
    & &   C$=$C               &   6.24              \\
    & &   C$=$N                &   6.37             \\
    & &   C$=$O               &   8.28             \\[0.05cm]
    & &   C$\equiv$C        &   8.65             \\
    & &   C$\equiv$N        &   9.19             \\[0.05cm]
\hline
\hline
                                & &                              &                   \\[-0.25cm]
\end{tabular}
\end{center}
\label{table_Si_bondE}
\end{table}

\subsection{UV photo-processing and the un-doping of a-C:H:Si} 
\label{sect_mechanism}

During the UV-EUV photolysis of a-C:H dust, particularly in PDRs, the Si atoms will be liberated from their $sp^3$ doping sites during the $sp^3 \rightarrow sp^2$ aromatisation process \citep[{\it e.g.},][]{2012A&A...540A...1J,2012A&A...540A...2J,2012A&A...542A...98J}. During this transformation to a trigonally-bonded C atom structure the growing (planar) aromatic domains will no longer be able to accommodate the tetrahedrally-bonded Si atoms and they will therefore be ejected into the gas phase.  The observed abundant gas phase Si  in PDRs, reflection nebul\ae\ and H{\footnotesize II} regions could therefore be a tracer of the UV photo-processing of a-C:H:Si materials accreted in dense regions of the ISM or formed in the dust shells around evolved stars. In the case of the abundant SiO observed in Herbig Haro jets the gas phase chemistry must rapidly oxidise liberated Si atoms to SiO. 
 
The SiC$_2$ molecule has been observed around carbon stars and in particular, and rather unexpectedly, in the inner dust-forming regions of \object{IRC\,+\,10216} \citep{2010A&A...521L...8C}. However, to date, SiC$_2$ has not been observed in the ISM. Thus, it would appear that any Si accreted into a-C:H mantles is not liberated as SiC$_2$ in PDRs  because the Si$_n$C$_m$ fragments must be rapidly photo-dissociated. 

Seemingly, PDRs are ideal sites for unraveling the a-C:H:$X$ grain composition via their EUV-UV photolytic deconstruction.

\section{O-doping}
\label{sect_O_doping}

The incorporation of O atoms does not appear to significantly affect the optical band gap of a-C:H materials but it does lead to a slight decrease in the refractive index, which goes hand-in-hand with a decrease in the density. Raman spectroscopy reveals that the addition of oxygen favours the clustering of the six-fold aromatic rings and FTIR spectroscopy shows that the material contains both C$-$O and C$=$O bonds but there is seemingly no evidence of O$-$H bonds \citep[{\it e.g.},][]{2004JAP....96.5456A}. In addition to a carbonyl band at $5.88\,\mu$m oxygen appears to have a marked effect on the $3.4\,\mu$m CH$_n$ bands \citep{2002ApJ...568..448G}. The presence of O atoms during a-C:H formation can promote the polymeric to aromatic transformation \citep[{\it e.g.},][]{2004JAP....96.5456A} but can also favour diamond-like clustering \citep[{\it e.g.},][]{2002ApJ...568..448G}. In the ISM there appears to be little evidence for the O-doping of a-C:H dust \cite[{\it e.g.},][]{2002ApJS..138...75P,2005A&A...432..895D}.

\section{N-doping}
\label{sect_N_doping}

\begin{table*}
\caption{Blue photoluminescence bands, on top of the broad continua (top two entries in italic) in laboratory ta-C:H (a-C:H materials rich in diamond-like or tetrahedral $sp^3$ C atom bonding) at 80\,K, excited by 300\,nm photons \citep{2006TSF...515.1597P} and in the \object{Red Rectangle} \object{HD~44179} \citep{2004ApJ...606L..65V,2005IAUS..235P.234V,2005ApJ...619..368V,2005ApJ...633..262V}. The most prominent sharp bands in the ta-C:H and ta-C:N laboratory data are marked in boldface. }
\begin{center}
\begin{tabular}{lcccccc}
                           &          &           &               &          &            &                          \\[-0.35cm]
\hline
\hline
                          &          &             &             &          &               &                          \\[-0.25cm]
 & \multicolumn{2}{c}{ta-C:H bands}       & \multicolumn{2}{c}{ta-C:N bands}       & \multicolumn{2}{c}{HD~44179  bands}   \\[0.05cm]
 & [ eV ]     &  [ nm ] & [ eV ]     &  [ nm ] &     [ eV ]   & [ nm ]  \\[0.05cm]

\hline
             &          &           &           &                &                  &              \\[-0.25cm]
{\it continuum}           &  {\it 2.2--3.7}       & {\it 335--564}          &  {\it 2.2--3.7}      & {\it 335--564}         & {\it 2.6--3.5} & {\it 350--480} \\[0.1cm]
{\it peak/shoulder}     &  {\it 3.2/---}         & {\it 390/---}              &  {\it 3.2/---}         & {\it 390/---}              & {\it 3.3/3.2} & {\it 375/390} \\[0.2cm]
 bands$\downarrow$ &  {\bf 3.52--3.54} &  {\bf 350.2--352.2}  &        3.51--3.54  &        350.2--353.2   &       ---        &     ---     \\
                                  &  ---                     &   ---                         &         3.42--3.43 &         361.5--362.5  &       ---        &     ---     \\
                                  &  {\bf 3.29}           &  {\bf 376.9}              &       3.28--3.31    &       374.6--377.5  &     3.28     &    377.5         \\
 strong                       &   ---                     &   ---                         &  {\bf 3.23--3.24}  &  {\bf 382.7--383.9} &     ---     &    ---         \\
                                  &  {\bf 3.09--3.11}  &  {\bf 389.7--401.2}  &        3.08--3.11   &         389.7--402.5  &    3.16     &    392.9         \\
 weak                        &        3.01             &        411.0               &  ---                      &   ---                        &    3.04     &    407.7        \\
 weak                        &        2.94             &        421.7               &       2.94--2.99    &   414.7--421.7       &    2.93     &    423.7        \\
                                  &   ---                     &       ---                      &       2.80             &   442.8                  &    2.83     &    438.8       \\
                                  &   ---                     &       ---                      &       2.79             &   444.4                  &    2.75     &     451.6      \\
 strong                       & {\bf 2.63--2.64}   &  {\bf 469.6--471.4}  & {\bf 2.63--2.64}   &  {\bf 469.6--471.4} &   2.67     &      465.2      \\[0.05cm]
\hline
\hline
                    &          &             &          &          &                       &                   \\[-0.25cm]
\end{tabular}
\end{center}
\label{table_blue_PL}
\end{table*}

The incorporation of N atoms into a-C:H aromatic structures, as C$-$N and C$=$N, promotes $sp^2$ carbon cluster formation, increasing the $sp^2/sp^3$ ratio as the hydrogen content decreases \citep[{\it e.g.},][]{1997PhRvB..5513020L,2001JAP....89.7924H,2004PhRvB..69d5410Y}. As the incorporated N fraction in an a-C:H network increases, the N atom bonding changes from  $\sigma$ bonding to three C atoms to $\pi$ bonding to two C atoms \citep[{\it e.g.},][]{2001JAP....89.7924H}. At low N content a-C:H materials retain their $sp^3$ diamond-like properties but as the nitrogen fraction increases, polymeric materials are formed, which include $>$C$-$N$-$ and also network-terminating C$\equiv$N and N$-$H groups that lead to decreased connectivity \citep{1999PSSAR.174...25R}. The presence of NH$_3$ and O$_2$ during formation has been shown to favour the formation of diamond-like crystallites in a-C:H \citep{2002ApJ...568..448G}. It has also been found that N-doping leads to a breaking of the symmetry and an activation of the olefinic C$=$C stretching mode \citep{1997PhRvB..5513020L} and decreases the optical gap by about 0.5\,eV \citep{1994DRM.....3.1034S,2001JAP....89.7924H}. 

It appears that the N-doping of a-C:H films is rather inefficient \citep{1994DRM.....3.1034S}  and the observable effects of nitrogen in the structure can seemingly be removed by heating to $\approx 800$\,K \citep{2002ApJ...568..448G}. Low N-doped a-C:H materials (N $<1$ at.\%), with an optical gap of $\simeq 1$\,eV, are rich with dangling bonds ($5 \times 10^{20}$\,cm$^{-3}$) and exhibit conductivity in the valence band tail. However, with increased N, the resistivity decreases and the optical band gap shrinks towards zero for high N doping \citep{1991JAP....70.4958A}.

Interestingly, the incorporation of N heteroatoms into the aromatic domains of a-C:H leads to ferro-magnetic properties in these  materials \cite[{\it e.g.},][]{2012JAP...111e3922L}.  Thus, the N-doping of a-C:H materials could have interesting consequences for the origin of interstellar polarisation in carbonaceous materials. 

Thus, N heteroatoms could clearly be incorporated into an accreting a-C:H $sp^2$ phase but this can probably only occur where there is sufficient UV to ensure the formation of an H-poor, aromatic-rich accreting material, {\it i.e.}, N would be expected to be incorporated into accreting a-C:H at lower $A_V$ than for Si-doping. Despite the apparent lack of heteroatomic N in interstellar hydrocarbon dust \citep[{\it e.g.},][]{2002ApJS..138...75P,2005A&A...432..895D}, N atoms could be incorporated into aromatic clusters and difficult to observe because of the similarity of the CC and CN infrared band positions, making the spectral interpretation and their detection rather difficult \citep[{\it e.g.},][]{2004PhilTransRSocLondA..362.2477F}. Nevertheless, \cite{2002ApJ...568..448G} have shown that the presence of N$_2$ during a-C:H formation can lead to broad bands at $6.17$ and $8\,\mu$m. The addition of nitrogen (and oxygen) heteroatoms leads to only subtle spectral changes\footnote{These manifest as a broadening of the CH$_2$ and CH$_3$ bands and variations in their relative intensities \citep{2002ApJ...568..448G}.} and the detection of doped a-C:H from $3.4\,\mu$m band spectroscopy will therefore be difficult \citep{2002ApJ...568..448G}. Thus, it appears that the N-doping of a-C:H, with a C$-$N bond energy of $\sim 3.2$\,eV, could be a route to N atom depletion in the ISM.

\subsection{Blue photoluminescence from a-C:H and N-doped a-C:H}  
\label{sect_BL_taCHN}

The photoluminescence (PL) spectra of hydrogen and nitrogen incorporated tetrahedral amorphous carbon materials (ta-C:H and ta-C:N, respectively) show a strong peak in the $\sim 2.2-3.7$\,eV ($335-564$\,nm) region, which is enhanced at low temperature and exhibits a number of sub-peaks on top of the broad peak  \citep{2006TSF...515.1597P}. Interestingly, the experimentally-measured blue PL  occurs over the same energy range, and has the same overall form, as the blue luminescence observed in the \object{Red Rectangle} (RR) region by \cite{2004ApJ...606L..65V,2005IAUS..235P.234V,2005ApJ...619..368V,2005ApJ...633..262V}, which is noted for its extended red emission (ERE). 
The ERE is seemingly widespread in the diffuse ISM \citep{1998ApJ...498..522G} but the blue luminescence has to date only been observed in the RR and in several reflection nebul\ae\ \citep{2005ApJ...633..262V}.

Noticeably, most of the RR blue PL bands \citep{2005IAUS..235P.234V} coincide with those seen in the laboratory ta-C:H and ta-C:N blue PL data (see Table~\ref{table_blue_PL}). This coincidence seems to indicate that an $sp^3$-rich a-C:H material, with some N-doping, could explain the observed blue PL in the RR. 

Given that ta-C:H(:N) materials are $sp^3$-rich or diamond-like it is perhaps not surprising that some of the observed sharp features in the RR could be attributed to diamond dust \citep{1988Ap&SS.150..387D}. Nor is it surprising that N atoms could be an important dopant in the RR dust because they are known to be the primary dopant in (nano-)diamond materials \citep[{\it e.g.},][]{2004ASPC..309..589J}.  Further, the steep rise in the FUV extinction in the RR found by \cite{2005ApJ...619..368V} is consistent with the optical properties of H-rich, wide band gap a-C:H materials derived by \cite{2012A&A...540A...1J,2012A&A...540A...2J,2012A&A...542A...98J}.

If the blue PL is associated with N-doped a-C:H grains, similar to an $sp^3$-rich ta-C:H:N material, it will be susceptible to aromatisation by EUV-UV photons \citep[{\it e.g.},][]{2012A&A...540A...1J,2012A&A...540A...2J,2012A&A...542A...98J}. In the RR it has been found that the blue PL is more extended than the ERE \citep{2005ApJ...619..368V}.  It would seem logical that the ERE PL is excited by lower energy photons than the higher-energy blue PL.  Thus, a rather natural interpretation of the observed PL is that the blue PL carrier is excited by UV photons from the ambient ISRF and that the ERE, which is found closer to the relatively cool central star \object{HD~44179} ($T_{\rm eff} = 8000$\,K), is excited by the lower-energy photons from the star. 

In the RR a number of the optical emission lines at 580.0, 585.3, 638.0 and 661.5\,nm, amongst others, bear a remarkable wavelength resemblance to several of the diffuse interstellar bands (DIBs) at 579.7, 585.0, 637.9 and 661.4\,nm \citep{1992MNRAS.255P..11S,1998MNRAS.301..955D}. Thus, it seems highly likely that DIB-carrying precursor species, if not the DIB carriers themselves, can be formed in the dust-forming regions around evolved stars.

\section{S-doping}
\label{sect_S_doping}

Commercial amorphous carbon doped with sulphur (a-C:S) has been shown to exhibit an inhomogeneous (type-II) superconductivity for $T < 38$\,K and to demonstrate an associated anomalous, spontaneous ferro-magnetic-like magnetisation \citep{2009PhRvB..79w3409F}. We again find that heteroatom doping can lead to unexpected magnetic properties in a-C:H materials, which could perhaps help to explain the weak UV 217\,nm bump polarisation observed along two lines of sight \citep{1997ApJ...478..395W}. 

The incorporation of S into accreting a-C:H mantles (at the $\approx 10$\% level\footnote{Given a sulphur abundance of $\cong 14$\,ppm and $140\pm20$\,ppm of carbon \citep{1996ApJ...467..334C} in the gas phase in the diffuse ISM.}) in the transition from the diffuse to the denser ISM could explain the disappearance of this element from the gas and its trapping into a difficult to observe form.   For example, the C$-$S the stretching band occurs in the $\simeq 15\,\mu$m region and C$=$S stretching in the $\sim 8-10\,\mu$m, both of which would be within the strong silicate absorption bands. It is possible that the S-doping of a-C:H, with a C$-$S bond energy of $\sim 2.8$\,eV, could therefore help to explain some of the broadening of the silicate absorption features but only in the denser (molecular) regions of the ISM where sulphur has significantly depleted from the gas.

\section{Mg-doping}
\label{sect_Mg_doping}

Given that the depletion of Mg in the diffuse ISM follows that of Si \citep[{\it e.g.},][]{2009ApJ...700.1299J} it is not unreasonable to assume that Mg would also incorporate into a-C(:H) as per Si (see Section \ref{sect_Si_doping}). Unfortunately there does not appear any relevant literature on the Mg-doping of a-C(:H) and so it is not yet possible to say anything about the relevance of a-C:H:Mg materials to astrophysics.    

\section{Fe-doping}
\label{sect_Fe_doping}

Relative to un-doped a-C:H materials Fe-doped hydrogenated amorphous carbons (a-C:H:Fe) are more aromatic, the Tauc optical gap is narrower by 0.3\,eV, the PL peak shifts from 2.35\,eV to 1.95\,eV, the PL intensity is greatly enhanced, and a deep level emission peak around 2.04\,eV (609\,nm) is observed \citep{2010ApPhA..98..895Z}. 
However, given that almost all iron is locked into dust in the ISM \citep[{\it e.g.},][]{1996ARA&A..34..279S,2000JGR...10510257J,2009ApJ...700.1299J} the Fe-doping of a-C:H is unlikely to be important in the ISM.

\section{The broader astrophysical implications} 
\label{sect_implications}

As dust transits from the diffuse ISM to denser regions it is expected that hydrocarbon-rich mantles will accrete in the relatively low-density and low-extinction ISM \cite[{\it e.g.},][]{1990QJRAS..31..567J,2011A&A...530A..44J,2013A&A...xxx...xxx} and that this must occur before water ice mantles accrete in molecular clouds ($A_V \gtrsim 3$). It is therefore to be expected that the accreting a-C:H mantles will also incorporate other relatively abundant gas phase elements such as Si, N, O and S, with bond energies of the order of 3\,eV for all (except for O atoms with a C$-$O bond energy of 3.7\,eV).  If this is the case, then in what order and in what form do these other elements accrete? For example, N atoms should show a preference for incorporation into $sp^2$-rich carbons and therefore ought to accrete into aromatic-rich a-C mantles before the gas phase Si, with a preference for the $sp^3$ phase, accretes into aliphatic-rich a-C:H mantles. 

\begin{table}
\caption{Some IR band signatures of dopant heteroatoms in a-C:H:$X$.}
\begin{center}
\begin{tabular}{lclc}
                     &                                 &                                                 &   \\[-0.35cm]
\hline
\hline
                      &                                 &                                                &   \\[-0.35cm]
          &  $\approx$ wavelength      &                                            &           \\
 $X$   & [ $\mu$m ]                &   origin                            &  note \\[0.05cm]
\hline
                     &                                   &                                                 &    \\[-0.25cm]
 Si     &   11.3                         &  \ -Si$-$C-                          &  a \\
         &   8.0, 12.7                  &  $>$Si(CH$_3$)$_2$              &  a \\
         &   4.4$-$4.7                 &  \ -Si$-$H                                    &  b \\[0.05cm]
  O     &   5.9                          &  $>$C$=$O carbonyl              &  b  \\
         &   3.1                           &  \ -O$-$H                               &  b \\[0.05cm]
  N     &   6.2, 8.0                    &   \ \ \ \ ?                             &  a \\
         &   4.4$-$4.5                 &   \ -C$\equiv$N nitryl             &  c \\
         &   4.8                           &   \ -N$\equiv$C isonitryl         &  c \\
         &   2.9$-$3.0, 6.3          &  $>$N$-$H,  -NH$_2$    &  a \\[0.05cm]
  S     &    15                           &  \ -C$-$S-                         &  a \\
         &    8$-$10                    &  $>$C$=$S                            &   a \\[0.05cm]
\hline
\hline
                     &                                   &                                                 &    \\[-0.25cm]
\end{tabular}
\begin{list}{}{}
Notes:
\item[] a. Possible confusion with amorphous silicate and/or aromatic hydrocarbon emission and absorption bands. 
\item[] b. There is currently no evidence for these absorption bands in the diffuse ISM \cite[{\it e.g.},][]{2002ApJS..138...75P,2005A&A...432..895D}. 
\item[] c. Bands not present in nitrogen-doped amorphous silicates heated to $\approx 800$\,K \citep{2002ApJ...568..448G}. 
\end{list}
\end{center}
\label{table_IR_bands}
\end{table}

The threads drawn together here seem to indicate that a-C:H grains and mantles can clearly be doped with Si, N, O and S atoms. This doping leads to some interesting consequences that could perhaps explain the `volatile' Si problem,  the origin of the blue luminescence, the sequestration of S and N atoms from the gas in the denser ISM and the magnetic behaviour of carbonaceous dust. Clearly, these same elements could also be incorporated into a-C:H in the dust-forming shells around evolved stars with equally interesting consequences. 

The heteroatom doping of a-C:H does lead to telltale IR signatures that ought to indicate their presence within the a-C:H:$X$ network. However, given their IR band positions it would appear that it will be difficult to unambiguously observe a-C:H dopants directly in the ISM. As shown in Table~\ref{table_IR_bands} most of the IR signature bands characteristic of heteroatom doping are coincident with the much stronger amorphous silicate absorption and aromatic emission bands in the ISM. In principle O heteroatoms could be observed indirectly through the changes that they induce in the shape and relative intensities of the component CH$_n$ bands ($n = 1,$ 2 or 3) in the $3.4-3.5\,\mu$m wavelength region but these effects are subtle \citep{2002ApJ...568..448G} and will be hard to disentangle from the intrinsic compositional variations in a-C(:H) materials \citep[{\it e.g.},][]{2012A&A...540A...2J,2012A&A...542A...98J}. 

The most viable carrier for the observed circumstellar and interstellar ERE is a relatively H-rich, wide band gap a-C:H material that can be excited over a wide wavelength range \citep[{\it e.g.}, $\lambda \gtrsim 250$\,nm,][]{2010A&A...519A..39G}. An N-doped a-C:H material could perhaps also provide an explanation for the observed blue PL and the superimposed, narrower features.   Apparently, this ERE-carrying $sp^3$, aliphatic-rich interstellar a-C:H dust component is only present in the cores of large ($a \sim 100-300$\,nm) carbonaceous grains \citep{2012A&A...540A...1J,2012A&A...540A...2J,2012A&A...542A...98J,2013A&A...xxx...xxx}.  These materials absorb strongly in the UV but their UV photon-processed, H-poor, a-C mantles absorb strongly in the visible and could thus pump the ERE excitation by the absorption of a wider range of photon energies than the wide band gap, H-rich a-C:H cores.  

In general, the Si in doped a-C:H will have a strong preference for the $sp^3$ phase and therefore tend to be ejected from the solid during UV photo-processing to $sp^2$-rich materials.  However, Si atoms could be incorporated into an $sp^2$ phase because the tetrahedral coordination of the Si atoms will allow for SiC$_4$ pentagon formation\footnote{This is because the tetrahedral bond angle (109.5$^\circ$) is close to the pentagon bond angle (108$^\circ$) but much smaller than the hexagonal or aromatic bond angle of 120$^\circ$.} and could play a role in fullerene-type cage formation by introducing curvature into an otherwise planar aromatic structure. Given that N heteroatoms show a preference for the aromatic $sp^2$ phase, it is likely that they can also be incorporated into fullerenes and fullerene-precursor arophatic structures \citep[{\it e.g.},][]{2012ApJ...761...35M}. Thus, C$_{60}$ and other fullerenes in the ISM could be accompanied by species such as C$_{(60-n)}X_n$ where $n$ is a small number and $X$ are Si or N atom substitutions for C atoms in the aromatic structure. Similarly, Si and N atoms could be incorporated into arophatic structures.  Such heteroatom substitutions will likely result in colour-centres in a-C:H and fullerene(-like) particles that could explain some of the observed optical emission bands that lie close to diffuse interstellar band positions in the RR \citep[{\it e.g.},][]{1992MNRAS.255P..11S}. Thus, it would seemingly be worthwhile to investigate these kinds of species as possible carriers of DIBs in the ISM. Fullerene-like structures, being more stable against UV photo-processing than arophatics, ought to provide a resistant DIB carrier population, whereas the $X$-doped arophatics would offer a more transient DIB-carrier population with many more available transitions because of the larger number of potential isomers. 


\begin{acknowledgements} 
This research was, in part, made possible through the financial support of the Agence National de la Recherche (ANR) through the programs Cold Dust (ANR-07-BLAN-0364-01) and CIMMES (ANR-11-BS56-029-02).
\end{acknowledgements}


\bibliographystyle{aa} 
\bibliography{biblio_HAC} 





\listofobjects

\end{document}